\documentstyle[preprint,aps,epsf,floats]{revtex}

\newcommand{\beq}{\begin{equation}}
\newcommand{\eeq}{\end{equation}}
\newcommand{\bea}{\begin{eqnarray}}
\newcommand{\eea}{\end{eqnarray}}
\newcommand{\ben}{\begin{eqnarray*}}
\newcommand{\een}{\end{eqnarray*}}

\newcommand{\simlt}{\stackrel{<}{{}_\sim}}
\newcommand{\simgt}{\stackrel{>}{{}_\sim}}

\def\vec#1{{\bf #1}}

\tightenlines

\begin{document}
\draft

\title{A renormalized equation for the three-body system with short-range
       interactions}

\author{H.-W. Hammer\footnote{hammer@mps.ohio-state.edu} 
and Thomas Mehen\footnote{mehen@mps.ohio-state.edu}}

\address{Department of Physics, The Ohio State University, Columbus, OH 
43210}

\date{\today}  

\maketitle

\begin{abstract}

We study the three-body system with short-range interactions characterized by
an unnaturally large two-body scattering length.  We show that the off-shell
scattering amplitude is cutoff independent up to power corrections.  This
allows us to derive an exact renormalization group equation for the three-body
force.  We also obtain a renormalized equation for the off-shell scattering
amplitude.  This equation is invariant under discrete scale transformations.
The periodicity of the spectrum of bound states originally observed by Efimov
is a consequence of this symmetry. The functional dependence of the three-body
scattering length on the two-body scattering length can be obtained
analytically using the asymptotic solution to the integral equation.  An
analogous formula for the three-body recombination coefficient is also
obtained.  \end{abstract}

\bigskip

\thispagestyle{empty}

\newpage                   

Recently, there has been renewed interest in the three-body system with
short-range interactions 
\cite{BvK98,BHK98,BHV99,BHK00,BeG00,BFG00,BBH00,BlG00,Wil00}. 
From the effective field theory (EFT) perspective,
an understanding of this system is an important ingredient for a successful
description of few- and many-body systems in nuclear and atomic physics
\cite{EFT98,EFT99,Birareview,BEANE99}. Despite its simplicity, this system
exhibits many interesting features \cite{Tho35,Efi71,Phi68}. For example, when
the two-body scattering length $a_2$ approaches infinity, the three-body system
exhibits an infinite number of shallow bound states \cite{Efi71}.  The
equations describing three particles interacting via strong short-range
two-body forces have been known for a long time \cite{STM57}.  While these
equations are well behaved in some channels, they do not have a unique 
solution for spinless bosons or three nucleons with total spin $S=1/2$
\cite{Dan61,DaL63}. 

The renormalization of the EFT for the three-body system with short-range
interactions has been discussed in detail in Ref.~\cite{BHV99}. Here we give a
brief summary of the results. For nonrelativistic particles interacting via
short-range forces, the Lagrangian consists of a nonrelativistic kinetic term
and an infinite number of contact interactions with an increasing number of
derivatives.  For systems in which the two-body scattering length $a_2$ is much
larger than the characteristic range of the interactions, the leading two-body
contact interaction with no derivatives needs to be treated nonperturbatively 
\cite{KSW98,UvK99,Geg98}. For three-body systems, the leading order Feynman
diagrams can be summed using an integral equation which is identical to that
of Ref.~\cite{STM57} if the three-body force is of natural size and subleading.
For three bosons or for the spin-1/2 state of three nucleons, this integral
equation exhibits strong dependence on the cutoff used to regulate the theory.
From the viewpoint of effective field theory, this is surprising as the
Feynman diagrams that the integral equation is designed to sum are individually
finite.  Their sum, however, does not converge and this is the origin of the
cutoff dependence in the integral equation.

The cutoff dependence of the three-body integral equation is properly
interpreted via renormalization theory. In this case, the sensitivity to the
cutoff indicates that a three-body contact interaction, which one would regard
as subleading on the basis of naive dimensional analysis, is in fact leading
order. In Ref.~\cite{BHV99}, it was shown that the cutoff dependence in 
the three-body equation can be properly renormalized with the 
inclusion of this three-body force.
The three-body force exhibits a very unusual renormalization group flow: it is
characterized by a limit cycle. The relevance of a single three-body operator
provides a compelling explanation for the existence of the Phillips line
\cite{Phi68}. The
effective theory has enjoyed successful phenomenological applications in
neutron-deuteron scattering \cite{BvK98,BHK98,BHK00,BeG00,BFG00}, 
the scattering of
$^4$He molecules \cite{BHV99}, and three-body recombination of atoms in
Bose-Einstein condensates \cite{BBH00}.

In this paper, we will show that once the three-body force is included, the
cutoff dependence of off-shell three-body amplitudes vanishes as the cutoff is
taken to infinity. This fact allows us to write down renormalized equations in
which the cutoff is completely removed.  These renormalized equations exhibit
invariance under discrete scale transformations.  This invariance is exact at
leading order in the effective theory and can be used to constrain the
functional form of three-body observables. The spacing of energy levels of
low-lying three-body bound states is a direct consequence of this symmetry. 
The invariance of the three-body observables under the discrete scale
transformations and the spacing of the energy levels has been derived 
previously in a very different manner by Efimov \cite{Efi71,Efi79}.  
We also show how asymptotic solutions
to the equations can be used to determine analytically the dependence of the
three-body scattering length and the three-body recombination rate on the
two-body scattering length.  The formula for the three-body scattering
length has also been derived in earlier work by Efimov \cite{Efi71,Efi79},
while the behavior of the three-body recombination rate has been extracted from
fits to numerical solutions of the equations \cite{BBH00}.  

We begin with the integral equation for the elastic scattering of a particle
and a bound state of the other two particles:
\begin{eqnarray}
\label{threebdy}
[F(p;k)]^{-1} K(k,p) = M(k,p;k;\Lambda) + {2\over \pi} \int_0^\Lambda 
dq\,  M(q,p;k;\Lambda) {\cal P}\bigg({q^2 \over q^2-k^2}\bigg) K(k,q) \, ,
\end{eqnarray}
where 
\begin{eqnarray}
F(p;k)&=&\frac{8}{3}\left({1\over a_2}+\sqrt{{3 p^2 \over 4} - m E_k}\right)\,,
\nonumber \\
M(q,p;k;\Lambda)&=&
{1\over 2pq} \ln\left|{q^2 +q p +p^2 -m E_k \over q^2 -q p +p^2 -m E_k}
\right| +{H(\Lambda) \over \Lambda^2}\,. \nonumber
\end{eqnarray}
Equation (\ref{threebdy}) has been derived in Ref.~\cite{BHV99}. Here $a_2$ is
the two-body scattering length, and the total energy is $m E_k = 3k^2/4
-1/a_2^2$.  $K(k,p)$ describes the S-wave scattering of a particle and a bound
state with momentum $\pm\vec{k}$ into a state with momentum $\pm\vec{p}$.  This
is an off-shell quantity except at $p=k$, where it is related to the
scattering phase shift, $k\cot\delta = 1/K(k,k)$.  The three-body scattering
length is simply $a_3 = -K(0,0)$.  The integral equation is regulated by the
ultraviolet cutoff $\Lambda$ and $H(\Lambda)/\Lambda^2$ is the contribution
from the three-body force. $H(\Lambda)$ depends on the three-body 
parameter, $\Lambda_*$, and evolves in such a way as to render the
solution insensitive to $\Lambda$.

We concentrate now on threshold scattering. Setting $k=0$ and defining $K(0,p)
\equiv a(p)$, we obtain the equation:
\begin{eqnarray}
\label{ap}
[F(p;0)]^{-1} a(p) &= & {1\over p^2+1/a_2^2}+{H(\Lambda) \over \Lambda^2}
+ {2\over \pi}\int_0^\Lambda dq\, M(q,p;0;\Lambda)\, a(q) \, .
\end{eqnarray}
It is clear from the numerical solutions of Eq.~(\ref{ap}) in Ref.~\cite{BHV99}
that $a(p)$ is independent of $\Lambda$ up to power suppressed corrections. 
This fact is crucial for the derivation of the renormalized equation.  On-shell
quantities such $K(k,k)$ are necessarily cutoff independent if the theory is
properly renormalized. However, $a(p)$ is an off-shell quantity for $p\neq 0$,
so it is not obvious that $a(p)$ should be cutoff independent as well. To see
why this is the case we will derive a renormalization group equation (RGE) for
$a(p)$.  We begin by noting that it is possible to derive an equation in which
the three-body force is eliminated by taking the difference of Eq.~(\ref{ap})
with $p \neq 0$ and the same equation with $p = 0$, 
\begin{eqnarray}
\label{ap2}
[F(p;0)]^{-1} a(p) &-&{3 a_2 \over 16} a(0) =  {1\over p^2+1/a_2^2} -a_2^2  \\
&& +  {2\over \pi} \int_0^\Lambda dq \left[ {1\over 2pq} 
\ln\left|{q^2 +q p +p^2 +1/a_2^2 \over q^2 -q p +p^2 +1/a_2^2} \right| 
-{1 \over q^2 +1/a_2^2} 
\right] a(q)\, . \nonumber
\end{eqnarray}
Acting on Eq.~(\ref{ap2}) with $\Lambda\, d/d\Lambda$, and  
using the fact that $a(0)$ must be cutoff independent, we derive the 
following RGE for $a(p)$:
\begin{eqnarray}
\label{apRGE}
[F(p;0)]^{-1} \Lambda {d \over d \Lambda} a(p) &=&
{2 \over \pi} 
\left[ {1\over 2p} \ln\left|{\Lambda^2 +\Lambda p +p^2 +1/a_2^2 \over 
\Lambda^2 -\Lambda p +p^2 +1/a_2^2} \right| 
-  {\Lambda \over \Lambda^2 +1/a_2^2}  \right] a(\Lambda) \\
&& +  {2\over \pi} \int_0^\Lambda dq \left[ {1\over 2pq}\ln\left|
{q^2 +q p +p^2 +1/a_2^2 \over q^2 -q p +p^2 +1/a_2^2} \right| -  
{1 \over q^2 +1/a_2^2} \right] \Lambda {d \over d \Lambda} a(q)\, . \nonumber
\end{eqnarray}
The first term on the right hand side of Eq.~(\ref{apRGE}) is of order
$a(\Lambda)\, Q^2/\Lambda^3$, where $\Lambda\gg Q=p,1/a_2$.
In the limit of large $\Lambda$, Eq.~(\ref{apRGE}) becomes a homogeneous 
integral equation:
\begin{eqnarray}\label{aprge}
[F(p;0)]^{-1} \Lambda {d \over d \Lambda} a(p) = 
{2\over \pi} \int_0^\Lambda dq \left[ {1\over 2pq}\ln\left|
{q^2 +q p +p^2 +1/a_2^2 \over q^2 -q p +p^2 +1/a_2^2} \right| -  
{1 \over q^2 +1/a_2^2} \right] \Lambda {d \over d \Lambda} a(q)\, , 
\end{eqnarray}
which is trivially solved by
\begin{eqnarray}\label{rgesoln}
\Lambda {d \over d \Lambda} a(p) = 0.
\end{eqnarray}
Thus Eq.~(\ref{aprge}) provides an explanation for the cutoff independence of
$a(p)$. It is conceivable that Eq.~(\ref{aprge}) has other solutions, however,
the numerical calculations of Ref.~\cite{BHV99}(cf.\ Fig.~6) show that 
Eq.~(\ref{rgesoln}) is the correct solution up to corrections that vanish as 
$\Lambda$ goes to infinity. We have not yet obtained an analytic proof of 
Eq.~(\ref{rgesoln}).

\begin{figure}[!t]
  \centerline{\epsfysize=9.0truecm \epsfbox{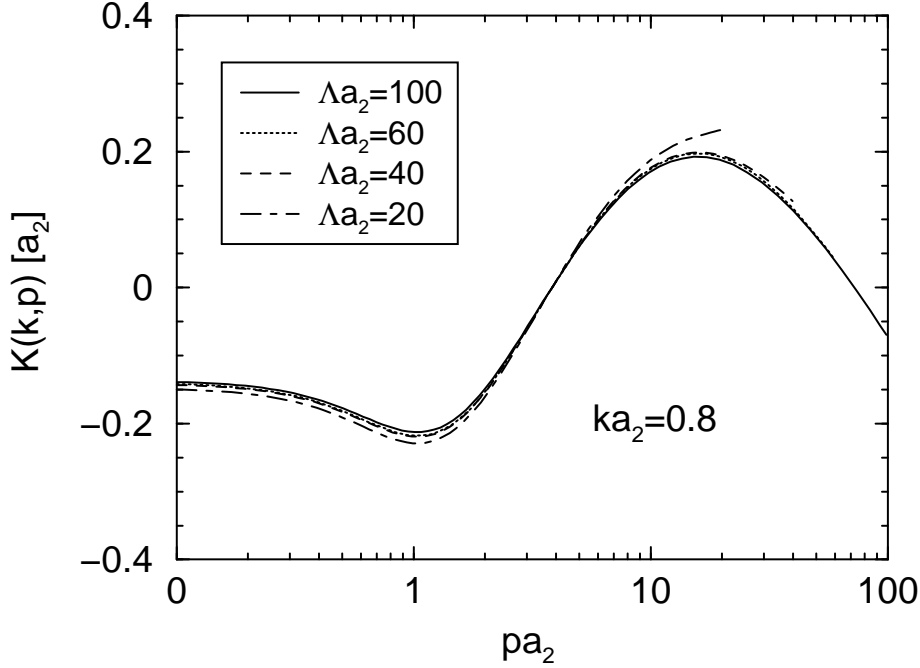}  }
 {\tighten
\caption[1]{The function $K(k,p)$ at $ka_2=0.8$ for $a_2 \Lambda_*=16$ 
 and four different values of the cutoff $\Lambda$.} 
\label{Kkp} }
\end{figure}

It is possible to repeat this analysis for the off-shell amplitude $K(k,p)$.
After eliminating $H(\Lambda)$ from Eq.~(\ref{threebdy}), acting with 
$\Lambda d/d\Lambda$, and dropping power suppressed terms, we find
\begin{eqnarray}
\label{Kkprge}
[F(p;k)]^{-1} &&\Lambda {d \over d \Lambda} K(k,p) =
[F(p;0)]^{-1} \Lambda {d \over d \Lambda} K(k,0)  \\
&&+ {2\over \pi} \int_0^\Lambda dq \left[ {1\over 2pq}\ln\left|
{q^2 +q p +p^2 -m E_k \over q^2 -q p +p^2 - m E_k} \right| -  
{1 \over q^2 - m E_k} \right] {\cal P}\bigg({q^2 \over q^2 - k^2 }\bigg) 
\Lambda {d \over d \Lambda} K(k,q)\, . 
\nonumber
\end{eqnarray}
Note that $K(k,p)$ is not symmetric so $K(k,0) \neq a(k)$. 
Eq.~(\ref{Kkprge}) is solved by
\begin{eqnarray}\label{Ksoln}
\Lambda {d \over d \Lambda} K(k,p) = 0.
\end{eqnarray}
In Fig.~\ref{Kkp}, we have plotted $K(0.8/a_2,p)$ for four different values of
the cutoff and $a_2 \Lambda_*=16$.  The function is cutoff
independent up to corrections of ${\cal O}(1/(\Lambda a_2 ))$. 
For the lowest value of the cutoff, $\Lambda a_2=20$, these corrections 
can be seen in Fig.~\ref{Kkp}. For the larger values of $\Lambda$,
they are negligible.

We can now use the cutoff independence of $a(p)$ to derive the RGE for the
three-body force.  Setting $p = 0$ in Eq.~(\ref{ap}) gives the following 
equation for the three-body scattering length $a_3$:
\begin{eqnarray}
\label{thbdysl}
-{3 a_2 \over 16} a_3= a_2^2 + {H(\Lambda) \over \Lambda^2}
+ {2 \over \pi}\int_0^\Lambda dq \left( {1 \over q^2 +1/a_2^2} +
{H(\Lambda) \over \Lambda^2} \right) a(q) \, ,
\end{eqnarray}
where we have used $a(0)=-a_3$.
All cutoff dependence in this expression is explicit. Taking derivatives
with respect to $\Lambda$ we obtain the exact RGE 
\begin{eqnarray}
\label{hrge}
\Lambda {d \over d \Lambda} \left({H(\Lambda) \over \Lambda^2} \right) =
-\frac{2 \Lambda}{\pi}\left(1+ \frac{2}{\pi} \int_0^\Lambda dq \, a(q)
\right)^{-1} \left( \frac{1}{\Lambda^2 +1/a_2^2}
+ \frac{H(\Lambda)}{\Lambda^2} \right) a(\Lambda)\, .
\end{eqnarray}
From a study of the asymptotics of Eq.~(\ref{ap}) 
(cf.\ Refs.~\cite{Dan61,BHV99}), we know the behavior of $a(p)$ for large $p$:
\begin{eqnarray}
\label{asymp}
a(p) = a_2 \,{\cal C}(a_2 \Lambda_*) \cos(s_0 \ln(p/\Lambda_*)) \, ,
\end{eqnarray} 
where $s_0\approx 1.006$ and 
$\Lambda_*$ is the three-body force parameter.
${\cal C}$ is an unknown dimensionless function of 
$a_2 \Lambda_*$ and expected to be of ${\cal O}(1)$. 
We can substitute this 
asymptotic solution for $a(p)$ into Eq.~(\ref{hrge}) in order 
to obtain an approximate RGE for $H(\Lambda)$. If we neglect terms of 
${\cal O}(1/(\Lambda a_2))$ we find 
\begin{eqnarray}
\label{rge_approx}
\Lambda {d \over d \Lambda}\left( {H(\Lambda) \over \Lambda^2} \right) =
{-(1+s_0^2)  \, \cos(s_0 {\rm ln}(\Lambda/\Lambda_*) ) \over 
\cos(s_0 {\rm ln}(\Lambda/\Lambda_*)+ s_0\sin(s_0 \ln(\Lambda/\Lambda_*)) } 
\left( \frac{1}{\Lambda^2}+ \frac{H(\Lambda)}{\Lambda^2} \right)\, ,
\end{eqnarray}
which is solved by 
\begin{eqnarray}
\label{H}
H(\Lambda) = - {\sin(s_0 {\rm ln}(\Lambda/\Lambda_*) -\arctan(1/s_0) )
\over \sin(s_0 {\rm ln}(\Lambda/\Lambda_*) +\arctan(1/s_0) )} \, .
\end{eqnarray}
This solution for $H(\Lambda)$ was previously obtained in Ref.~\cite{BHV99}
by requiring the low-energy solution $a(p\sim 1/a_2)$ to be 
invariant under finite changes of the cutoff. Equations (\ref{rge_approx},
\ref{H}) receive corrections that scale as $1/(a_2 \Lambda)$, which could 
in principle be computed with the help of Eq.~(\ref{hrge}). In practice, 
Eq.~(\ref{H}) is an excellent approximation to the exact evolution of 
$H(\Lambda)$ (cf.\ Fig.~8 of Ref.~\cite{BHV99}). 

We are now in a position to write renormalized equations for 
$K(k,p)$ and  $a(p)$. Cutoff independence of $K(k,p)$ implies 
that we are free to choose the cutoff to be whatever we like.
Up to corrections of ${\cal O}( 1/(a_2 \Lambda))$,
$H(\Lambda)$ vanishes if $\Lambda =\Lambda_n$ where
\begin{eqnarray}
\label{zero}
\Lambda_n &=& \Lambda_* \exp\left[\frac{1}{s_0}\left(n \pi + 
\arctan\left(\frac{1}{s_0}\right) \right) \right]
\approx \Lambda_* \exp\left[\left(n+\frac{1}{4}\right)\pi\right]
\, .
\end{eqnarray}
Setting $\Lambda =\Lambda_n$ in Eqs.~(\ref{threebdy},\ref{ap}) results in
\begin{eqnarray}\label{renKkp}
[F(p;0)]^{-1}K(k,p) &=& {1\over 2pk} \ln\left|{k^2 + k p +p^2 -m E 
\over k^2 -k p +p^2 -m E}\right| \\
&& + {2\over \pi} \int_0^{\Lambda_n (\Lambda_* )} dq {1\over2 pq} 
\ln\left|{q^2 +q p +p^2 - m E \over q^2 -q p +p^2 -m E} \right| \, 
{\cal P}\bigg({q^2 \over q^2-k^2}\bigg) K(k,q) \, , \nonumber
\end{eqnarray}
and
\begin{eqnarray}
\label{reneq}
[F(p;0)]^{-1}a(p) = {1\over p^2+1/a_2^2} 
+ {2\over \pi} \int_0^{\Lambda_n (\Lambda_* )} dq {1\over2 pq} 
\ln\left|{q^2 +q p +p^2 +1/a_2^2 \over q^2 -q p +p^2 +1/a_2^2} \right| \, 
a(q) \, .
\end{eqnarray}
These are interesting equations because all dependence on $\Lambda$ has been
removed in favor of the physical parameter characterizing the three-body
force, $\Lambda_*$, which appears in the upper limit of the integral.
These equations receive corrections which fall off as
powers of $\Lambda$. The leading corrections in $1/\Lambda$ come from the
running of $H(\Lambda)$ and scale as $1/(a_2 \Lambda)$. They are suppressed 
in the limit $a_2 \rightarrow \infty$
and can be made negligible by choosing $n$ sufficiently
large.  Note that we are not taking the cutoff to infinity in a
continuous manner, rather in a series of discrete steps for which
$H(\Lambda) = 0$.  

The renormalized equations have an exact symmetry that is a consequence of 
the freedom to choose an arbitrary integer value for $n$ in Eq.~(\ref{zero}).
This corresponds to rescaling $\Lambda_*$ by a factor of $\exp(n\pi/s_0)$
and holding all other dimensionful quantities fixed. Note that this symmetry 
is trivially satisfied in the two-body system. As a consequence,
physical observables should not change if the three-body force parameter is 
rescaled by $\exp(n\pi/s_0)$. 

This symmetry can also be seen in the unrenormalized equation. The amplitude
$a(p)$ is a function of three variables: $a(p;a_2,\Lambda_*)$. This function
has a remnant of scale invariance due to the periodicity of $H(\Lambda)$. 
If we perform the scale transformations:
\bea
\label{strafo}
p &\to& \alpha p\,, \\
1/a_2 &\to& \alpha/a_2\,,\nonumber\\
\Lambda &\to& \alpha \Lambda\,,\nonumber\\
a(p;a_2,\Lambda_*) &\to& \alpha^{-1} a(\alpha p;\alpha^{-1}a_2,\Lambda_*)
\nonumber
\eea
in Eq.~(\ref{ap}) where the cutoff is still present, we find
\bea
\label{rescaleeq}
& &[F(p;0)]^{-1} \,a(\alpha p;\alpha^{-1} a_2,\Lambda_*) = \\
& &\qquad\qquad {1\over p^2+1/a_2^2} + \frac{H(\alpha\Lambda)}{\Lambda^2}
+ {2\over \pi} \int_0^{\Lambda} dq \, M(q,p;0;\alpha\Lambda)\,
a(\alpha q;\alpha^{-1} a_2,\Lambda_*)\,. \nonumber
\eea
Note that all dimensionful quantities except for $\Lambda_*$ are
rescaled in the transformation in Eq.~(\ref{strafo}).
Up to corrections of ${\cal O}( 1/(a_2 \Lambda))$
(which can be made arbitrarily small
by choosing an appropriate cutoff $\Lambda$), 
the three-body force runs according to Eq.~(\ref{H}) and has a limit 
cycle with period $\exp(n\pi/s_0)$.
As a consequence, if $a(p;a_2,\Lambda_*)$ is a solution,
then $a(\alpha p, \alpha^{-1} a_2, \Lambda_* )$ is a solution as well 
for all $\alpha=\exp(n\pi/s_0)$. 

This symmetry leads immediately to the unique spectrum of three-body bound
states originally discovered by Efimov \cite{Efi71}. The equation for the
three-body bound state can easily be obtained from Eq.~(\ref{ap}). Since the
bound state solutions correspond to standing waves, we drop the inhomogeneous
term and replace $E_k$ with $-B_3$.  It is sufficient to consider the equation
in the limit $k=0$, and we have
\beq
\label{bound}
{\tilde a}(p)=\frac{16}{3\pi}\left({1\over a_2} + \sqrt{{3 p^2 \over 4} 
+ m B_3 }\right) \int_0^{\Lambda} dq \,\left[ 
\frac{1}{2pq}\ln\left|{q^2 +q p +p^2 +m B_3 \over q^2 -q p +p^2 
+m B_3} \right|+\frac{H(\Lambda)}{\Lambda^2} \right]\,{\tilde a}(q)\,,
\eeq
where ${\tilde a}(p)$ is related to the bound state wave function.  The binding
energies are then given by those values of $B_3$ for which Eq.~(\ref{bound})
has nontrivial solutions.  Equation (\ref{bound}) has the same symmetry as
Eq.~(\ref{ap}) if we transform $B_3 \to \alpha^2 B_3$. Consequently, if
Eq.~(\ref{bound}) has a solution for one value of $B_3\sim 1/(m a_2^2)$, the
symmetry then implies the existence of a spectrum of bound states with binding
energies \cite{Efi71}
\beq
\label{efi1}
B_3^n = d(s_0 \ln(a_2 \Lambda_*)) {\exp(2n\pi/s_0)\over m a_2^2}\,,
\eeq
where $n >0$ is an integer. The prefactor $d(s_0 \ln(a_2 \Lambda_*))$ 
is a periodic function of order one and has been calculated numerically
in Ref.~\cite{BHV99}. Because a stable three-body bound state
cannot have a binding energy
smaller than the two-body binding energy $B_2=1/(m a_2^2)$, $n$ has to be
larger than zero. Furthermore, due to the presence of the momentum cutoff
$\Lambda$ the maximal binding energy is of order $\Lambda^2/m$.  Inverting
Eq.~(\ref{efi1}), we then find for the total number of bound 
states \cite{Efi71}
\beq
N=\frac{s_0}{\pi} \ln(\Lambda a_2)\,.
\eeq
If we increase the cutoff, the maximal binding energy increases as
well\cite{Tho35}.
The shallow bound states, however, are kept at constant binding
energy through the running of $H$ with the cutoff
$\Lambda$ \cite{BHV99}. Corrections to Eq.~(\ref{efi1}) from
irrelevant operators have been studied in Ref.~\cite{Wil00}.

The discrete scale symmetry plays an important role in constraining the
functional dependence of other three-body observables. We will consider the
three-body scattering length $a_3$, the three-body effective range parameter
$r_3$, and the recombination coefficient $\alpha$.  In the recombination
process, three atoms start at rest, then two of the atoms form a bound state,
and the third atom balances energy and momentum. The recombination rate for
three atoms in a gas of density $n$ is $\nu = \alpha n^3$, where $\alpha$ is
the matrix element squared for the transition to take place. The integral
equation that determines $\alpha$ is very similar to that of $a(p)$ and will be
given below.  The quantities $a_3$, $r_3$ and $\alpha$ depend only on $a_2$ and
$\Lambda_*$ and the requirement that observables be invariant under $\Lambda_*
\rightarrow \exp(\pi/s_0) \Lambda_*$ highly restricts their functional form:
\begin{eqnarray}
\label{fform}
a_3(a_2,\Lambda_*) = a_2\, f(s_0 \ln(a_2 \Lambda_*)) \, , \\
r_3(a_2,\Lambda_*) = a_2\, g(s_0 \ln(a_2 \Lambda_*)) \, , \nonumber \\
\alpha(a_2,\Lambda_*) = \frac{a_2^4}{m}\, h(s_0 \ln(a_2 \Lambda_*)) \, , 
\nonumber 
\end{eqnarray}
where $f(x), g(x)$ and $h(x)$ are all periodic functions: $f(x)=f(x+\pi)$. 
The dependence on factors of $a_2$ and $m$ is determined by dimensional 
analysis.

In the following, we will investigate the observables in Eq.~(\ref{fform}) in
more detail. It turns out that the functional form of $a_3$ and $\alpha$ can be
obtained from the asymptotic solution. We first turn to the three-body
recombination rate $\alpha$. For three atoms approximately at rest in the
initial state, the final momentum for the bound state is $p_f=2/({\sqrt 3}a_2
)$. In Ref.~\cite{BBH00}, it was shown that $\alpha = |b(p_f)|^2$, where $b(p)$
is the solution of
\bea
\label{bp}
b(p)&=&\frac{32\pi\sqrt{2}}{\sqrt{m}\sqrt[4]{3}}\left(\frac{1}{p^2}+
     \frac{H(\Lambda)}{\Lambda^2}\right)\\
& &\qquad
+\frac{2}{\pi}\int_0^\Lambda \frac{dq\,q^2 b(q)}{-1/a_2 +\sqrt{3}q/2-i\epsilon}
  \left[\frac{1}{pq}\ln\left|\frac{q^2+pq+p^2}{q^2-pq+p^2}\right|
  +\frac{H(\Lambda)}{\Lambda^2}\right]\,. \nonumber
\eea
It is straightforward to show that $p\,b(p)$ satisfies the same
asymptotic equation as $a(p)$ from Eqs.~(\ref{ap}, \ref{reneq}). Consequently,
the ultraviolet behavior of $a(p)$ and $p\,b(p)$ is the same and 
Eq.~(\ref{bp}) is renormalized by the same three-body force \cite{BBH00}.
Furthermore, $b(p)$ is independent of $\Lambda$ up to power corrections, and
we can choose the cutoff as in Eq.~(\ref{zero}) to obtain a renormalized
equation.

The asymptotic solution for $b(p)$ is expected to be valid only for 
$p \gg 1/a_2$. However, in the numerical calculations of Ref.~\cite{BHV99} 
it was observed that the asymptotic solution works surprisingly well down to 
momenta of order $1/a_2$. In this case we should be able to use the asymptotic 
solution to get the functional form of $\alpha$. The solution to the
asymptotic equation for $b(p)$ has the form:
\beq
b(p) \sim \frac{a_2}{\sqrt{m}}\frac{\cos(s_0 \ln(p/ \Lambda_*))}{p}\, ,
\eeq
where the
factors of $a_2$ and $m$ are determined from dimensional analysis.
Note that the asymptotic equation does not
determine the normalization of the solution. Evaluating the formula for 
$\alpha$, we obtain
\begin{eqnarray}
\alpha = |b(p_f)|^2 \propto \frac{a_2^4}{m} \cos^2[s_0\ln(a_2\Lambda_*)
-0.1]\,.
\end{eqnarray}
The recombination coefficient was computed numerically in Ref.~\cite{BBH00}.
The numerical solution can be fit with the following expression: 
\begin{eqnarray}
\label{approx_alpha}
\alpha \approx \frac{a_2^4}{m} \, 68 \cos^2 [s_0 \ln (a_2 \Lambda_*) 
+ 1.7] \, .
\end{eqnarray}
Equation (\ref{approx_alpha}) is invariant under $\Lambda_* \rightarrow
\exp(\pi/s_0)\Lambda_*$ as expected. The asymptotic solution gives the correct
functional form but the normalization is not predicted. It does not
predict the correct phase in Eq.~(\ref{approx_alpha}) because the phase
of the full solution at $p \sim 1/a_2$ is not equal to the asymptotic phase.

Now we turn to the three-body scattering length which is given by 
$a_3 = -a(0)$. For $p\simlt 1/a_2$, the asymptotic
form is not a good approximation anymore. In fact, it
does not even have a well defined limit as $p\to 0$.
The only extra piece of information we
need is the existence of momenta for which the value of $a(p)$ is independent
of $\Lambda_*$. These meeting points were first observed in Ref.~\cite{BHV99}
(cf.\ Fig.~5). The first of these points occurs at $p_0=1.1/a_2$,
and additional points occur at $p_n = p_0 \exp(n\pi/s_0)$ with $n\geq 0$ an
integer. The existence and location of
these points is not determined by the asymptotic equation, but from
matching to the solution for $p\sim 1/a_2$. While 
the meeting points occur at off-shell 
momenta and are not observable, they severely constrain
the form of the solution for $p \simgt 1/a_2$. Taking into account the meeting
points, the solution for momenta $p\simgt 1/a_2$ is
\cite{BHV99}
\beq
\label{asymp2}
a(p) \sim a_2\, {\cal C}'\, \frac{\cos(s_0 \ln(p/\Lambda_*))}{\cos(s_0 
     \ln(p_0/ \Lambda_*))}\,,
\eeq
where
${\cal C}'$ is a constant that is independent of $a_2 \Lambda_*$.

We now insert this form into Eq.~(\ref{reneq}), set $p=0$, and keep only the
terms that dominate in the limit $a_2 \to \infty$. We split the
integral over $q$ into an integral from 0 to $p_0$ and an integral
from $p_0$ to $\Lambda_n$. For large $a_2$ the first integral is
parametrically suppressed and can be neglected. The second integral
can be performed by expanding the kernel in a formal power series,
\bea \label{a3}
 \frac{a_3}{a_2}&=& -\frac{16}{3} + a_2\, {\cal C}'\frac{32}{3\pi}
\int_{p_0}^{\Lambda_n} dq \sum_{n=1}^{\infty} (a_2 q)^{-2n}
(-1)^{n+1} \frac{\cos(s_0 \ln(q/\Lambda_*))}{\cos(s_0 
     \ln(p_0/ \Lambda_*))}\,. \nonumber\\
&=&-\frac{16}{3} -\frac{32}{3\pi} {\cal C}'\sum_{n=1}^{\infty}
\frac{(-1)^{n+1} (p_0 a_2)^{1-2n}}{s_0^2+(2n-1)^2}
 \left[ (2n-1) +s_0 \tan(s_0 \ln
(a_2 \Lambda_*)-0.1) \right]\,.
\eea
In Eq.~(\ref{a3}), we have kept only the leading terms for
for large $a_2$. The three-body 
scattering length obtained from a fit to numerical solutions in 
Ref.~\cite{BBH00} is 
\begin{eqnarray}
\label{scl}
a_3 \approx a_2 (1.4 - 1.8 \, {\rm tan}[s_0(a_2 \Lambda_*) + 3.2]) \, .
\end{eqnarray}
Again this is invariant under $\Lambda_* \rightarrow \exp(\pi/s_0)\Lambda_*$. 
Efimov gave an argument for this functional form in Ref.~\cite{Efi79}. The 
formula obtained using Eq.~(\ref{asymp2}) has the same functional form as 
Eq.~(\ref{scl}) but does not predict the numbers in Eq.~(\ref{scl}). One can 
also consider the effect of corrections to the asymptotic equation of the form
\begin{eqnarray}
\frac{\cos(s_0 \ln(p/\Lambda_*))}{\cos(s_0 
     \ln(p_0/ \Lambda_*))} \frac{c_n}{(p a_2)^n}\,.
\end{eqnarray}
These are subleading in the limit of large $p$ but become important for $p\sim
1/a_2$. The coefficient $c_1$ was computed in Ref.~\cite{Dan61}.  Corrections
of this form change the coefficients appearing in Eq.~(\ref{a3}) but do not
change the functional form.

\begin{figure}[!t]
  \centerline{\epsfysize=9.0truecm \epsfbox{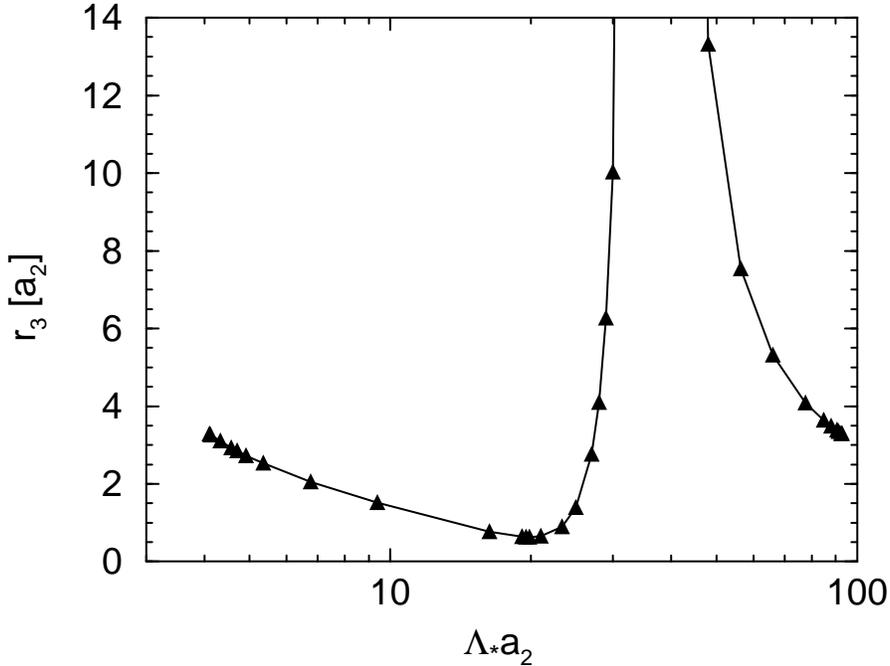}  }
 {\tighten
\caption[1]{Three-body effective range $r_3$ as a function of $a_2\Lambda_*$.} 
\label{r3} }
\end{figure}

It would be interesting to know the functional form of $r_3$. An analytic
expression for the effective range has not been derived. We have extracted
$r_3$ from numerical solutions to Eq.~(\ref{threebdy}). In Fig.~\ref{r3},
we have plotted $r_3/a_2$ as a function of $\Lambda_* a_2$ for
$\Lambda_* a_2$ between 4.1 and 92.9. 
These values of $\Lambda_* a_2$ differ by $22.7
\approx \exp(\pi/s_0)$, so this interval corresponds to one period
of the limit cycle, after which $r_3$ returns to its original value. 
In order to compute the functional form of $r_3$, it is necessary
to know the $k$ dependence of the amplitude $K(k,q)$. The asymptotic solution,
however, is independent of $k$. 

Knowing how three-body observables depend on the underlying parameters $a_2$
and $\Lambda_*$ is critical for understanding implications of nonrelativistic
scale and conformal symmetry.  In Ref.~{\cite{MSW00}, it was conjectured that
nonrelativistic systems interacting through short-range forces could be
invariant under nonrelativistic scale and conformal transformations in the
limit where $a_2$ goes to infinity and the two-body effective range parameters
are set to zero. This is interesting as it would give an example of a strongly
interacting system that is nevertheless scale and conformally invariant.  The
many-body problem in this limit has been suggested as a model for neutron
matter \cite{Bertsch} and could be relevant for gases of trapped atoms. In
Ref.~{\cite{MSW00}, it was shown that on- and off-shell two-body Green's
functions respect the nonrelativistic scale and conformal Ward identities. Ward
identities were also derived for three-body scattering.  Defining the 
limit $a_2\rightarrow \infty$ for the three-body system
is subtle because of the oscillations in
Eq.~(\ref{fform}). When we take $a_2$ to infinity continuously, $a_3$ will 
oscillate rapidly between $- \infty$ and $+\infty$. While this limit
is appropriate for atomic systems near Feshbach resonances \cite{BBH00},
other limits may be necessary to find scale and conformally
invariant theories.
One sensible limiting procedure for this purpose
would be to take $a_2 \rightarrow \infty$ in
discrete steps 
\begin{eqnarray}
\label{limit} \lim_{a_2 \rightarrow \infty}
\equiv \lim_{n \rightarrow \infty} (a_2 =\exp(n \pi/ s_0)a_2^0) \, ,
\end{eqnarray} 
where $n$ is a sequence of integers and $a_2^0$ is an arbitrarily
chosen starting point for the sequence of discrete transformations. An
alternative limiting procedure 
would be to take $a_2 \rightarrow \infty$ continuously while
taking $\Lambda_* \rightarrow 0$ in such a way that $a_2 \Lambda_*$ remains
constant.  When the $a_2 \rightarrow \infty$ limit is defined in either of
these two ways, observables of the form in Eq.~(\ref{fform}) live at one point
in the limit cycle. $a_3$ does not oscillate when the limit $n \rightarrow
\infty$ is taken but tends to either zero or $\pm \infty$, depending on the
initial choice $a_2^0$.  Obviously obtaining zero requires some fine tuning of
$a_2^0$ (or equivalently, $\Lambda_*$). $r_3$ will always tend to $+\infty$
because it is positive for all values of $\Lambda_*$ in its limit cycle, as is
seen in Fig.~\ref{r3}.  Higher order terms in the effective range expansion
will tend to either zero or $\pm \infty$.  Because these 
higher order terms can also diverge as $a_2 \rightarrow  \infty$,
one needs to understand the
behavior of the entire function $K(k,k)$ as $a_2 \rightarrow \infty$. It would
be interesting to determine this behavior and check if the three-body amplitude
obeys the scale and conformal Ward identities derived in Ref.~\cite{MSW00}.

One very useful property of the renormalized equations is that it is possible
to demonstrate analytically that the theory is being renormalized properly
rather than having to check this numerically with each calculation.  The
existence of renormalized equations will facilitate future numerical work, 
since there are no delicate cancellations between two different 
terms in the kernel. 
{\em We emphasize that we are not eliminating the three-body 
force but using the renormalization group to simplify the form of the 
equation and make the dependence on the three-body force parameter, 
$\Lambda_*$, explicit.}  In some
early treatments of the three-body problem \cite{Kar73}, the integral equation
was solved with a finite cutoff that was tuned to fit observed data and then
treated as a universal parameter. Though this procedure may seem somewhat ad
hoc, Eq.~(\ref{reneq}) shows that this is in fact a rigorous procedure.

For many applications it will be important to understand how to explicitly
renormalize higher orders in the EFT. For low-energy neutron-deuteron
scattering it is clear that effective range corrections are crucial for
obtaining accurate agreement with data 
\cite{BvK98,BHK98,BHK00,BeG00,BFG00}. Being able to renormalize
equations analytically at higher orders will facilitate these increasingly
complex calculations. It would be advantageous to understand the
renormalization group evolution of higher dimension three-body operators to see
if they evolve according to similar limit cycles and to understand at what
order these enter into calculations.

A renormalized equation for two-nucleon systems in theories with explicit pions
would also be of great practical value.  Pion exchange gives rise to a $1/r^3$
potential which must be treated nonperturbatively. It is not obvious that the
resummed pion graphs can be renormalized by a single local operator. If this is
possible, one would expect this operator to exhibit a limit cycle similar
to the one observed in the three-body system, and one would like to calculate
the evolution of this operator analytically. Investigations along these lines
have been carried out in a position space treatment using a nonrelativistic
Schr\"odinger equation regulated with square well potentials \cite{Bea00}.  

Interesting applications also include coupling photons and weak currents to few
nucleon systems. In these applications one would like to employ gauge invariant
regulators such as dimensional regularization. This is also important for
theories with pions where one would like to preserve chiral symmetry
explicitly. In this context, it would be useful to have renormalized
integral equations which are regulated using dimensional regularization.

We acknowledge useful discussions with P.F.\ Bedaque, E.\ Braaten, 
R.J.\ Furnstahl, and M.B. Wise.
This work was supported by the National Science Foundation
under Grant No.\ PHY--9800964.

\end{document}